\documentclass[reprint,superscriptaddress,secnumarabic,amssymb, nobibnotes, aps, pra]{revtex4-2}

\usepackage{amsmath} 
\usepackage{physics}
\usepackage{lineno,hyperref}
\usepackage{graphicx}
\usepackage{epsfig}
\usepackage{subfigure}

\newcommand{\be}{\begin{equation}}
\newcommand{\ee}{\end{equation}}
\newcommand{\bea}{\begin{eqnarray}}
\newcommand{\eea}{\end{eqnarray}}

\renewcommand{\vec}[1]{\boldsymbol{#1}} 
\begin{document}
\bibliographystyle{apsrev4-2}
\title{Polarons in a ferromagnetic spinor Bose-Einstein condensates}%
\author{Xiao-Lu Yu}%
\email{yuxiaolu1983@gmail.com}
\noaffiliation
\author{Boyang Liu}
\affiliation{Institute of Theoretical Physics, Beijing University of Technology Beijing 100124, China}
\date{\today}%
\begin{abstract}
We investigate the polarons formed by immersing a spinor impurity in a ferromagnetic state of $F=1$ spinor Bose-Einstein condensate. The ground state energies and effective masses of the polarons are calculated in both  weak-coupling regime and strong-coupling regime. In the weakly interacting regime the second order perturbation theory is performed. In the strong coupling regime we use a simple variational treatment. The analytical approximations to the energy and effective mass of the polarons are constructed. Especially, a transition from the mobile state to the self-trapping state of the polaron in the strong coupling regime is discussed. We also estimate the signatures of polaron effects in spinor BEC for the future experiments.
\end{abstract}
\maketitle
\section{Introduction}
The polaron is a state formed in a system of a single distinguishable particle interacting with medium atoms, which can be used to describe various systems in traditional solid states \cite{doi:10.1080/00018735400101213,mahan1990many}, quantum liquids \cite{PhysRev.127.1452,RevModPhys.63.675}, advanced materials \cite{Alexandrov2007PolaronsIA}, chemistry\cite{emin_2012,doi:10.1021/ar00118a005} and biophysics \cite{SCOTT19921,Conwell8795}. Recently, an impurity atom immersed in a Bose-Einstein condensate (BEC) has been experimentally realized \cite{PhysRevLett.117.055301,PhysRevLett.117.055302},
and there are intensive interests focusing on the physics of Bose polarons\cite{PhysRevLett.93.120404,:80302,PhysRevB.80.184504,PhysRevLett.96.210401,PhysRevX.8.011024,PhysRevLett.123.213401,PhysRevA.88.053632,PhysRevA.90.013618,PhysRevA.89.053617,PhysRevA.88.053610,Ardila_2018,PhysRevLett.120.050405,PhysRevLett.123.213401,PhysRevA.100.043605,PhysRevA.101.013623}. The main motivations of these simulations are to explore the hitherto inaccessible strong coupling regime in polaron physics by utilizing the remarkable tunabilities of experimental techniques in the ultracold gas systems \cite{RevModPhys.82.1225}.

The purpose of the present paper is to generalize the polaron effects in a spinor impurity-spinor BEC system. The spin of the alkali atoms is essentially free in an optical trap, producing a rich variety of spin textures. The spin-polaron effects arise from spin dependent interactions between an impurity and BEC atoms, which has several features distinct from the scalar BEC. First, the scalar BEC systems only provide a gapless phonon mode. There are many collective modes in a spinor BEC, which could be gapless or gapped, coupling with the impurity to induce polaron effects. This properly provides the scenario to simultaneously simulate optical and acoustic polaron of solid-states \cite{doi:10.1080/00018735400101213,10.1143/PTP.26.29,1973137} with spinor quantum gases. Second, the energies and effective masses of polarons among different spin components will split, due to spin-exchange collisions between an impurity and BEC atoms. These are characteristic phenomena in spin-polaron physics, which plays an important role in strongly correlated electron materials \cite{PhysRev.118.141,Alexandrov2007PolaronsIA}. Third, with the techniques of optical Feshbach resonance \cite{PhysRevLett.114.255301}, it is possible to manipulate polaron with different spin components, such as the formation of self-trapping states in the strong-coupling regime.

The method of the present paper is to map the spinor impurity-BEC system onto Fr\"{o}hlich Hamiltonian of original polaron problem, then analogous effects are derived in quantum gas systems. Fr\"{o}hlich Hamiltonian is one of the simplest examples of a quantum field theoretical problem, since it basically consists of a single particle interacting with a Bose field. There are many variations of  Fr\"{o}hlich Hamiltonian, which are studied thoroughly in \cite{Devreese_2009, MITRA198791}. However, a spinor generalization of  Fr\"{o}hlich Hamiltonian has not been explicitly spelled out in any real physical system to our knowledge, which is what we attempt to do here. 

The paper is organized as follows. The spinor generalization of Fr\"{o}hlich polaron Hamiltonian is described in Sec. \ref{polH}.  In the weak-coupling regime, we illustrates the polaron arising from spin-dependent interactions by a perturbation theory in Sec. \ref{weak}. With a simple variational treatment in Sec. \ref{strong}, we construct analytical approximations to the energy and effective mass of polaron in the strong-coupling regime. The splitting of energy and effective mass among different spin components are emphasized. Sec. \ref{exp} contains experimental discussions and conclusions.
\section{Polaron Hamiltonian}\label{polH}
Let us consider a system of homogenous spin-1 Bose gas with s-wave interaction. This system can be described in the second quantized Hamiltonian \cite{PhysRevLett.81.742} as
\begin{equation}
\begin{aligned}
H_{BEC}=&\sum_{\vec{k},f}\epsilon_{\vec{k}}a_{\vec{k},f}^{\dag}a_{\vec{k},f}+\frac{1}{2}\sum_{\vec{k},f}\delta_{\vec{k}_{1}+\vec{k}_{2},\vec{k}_{3}+\vec{k}_{4}}\\
&\times(c_{0}a_{\vec{k}_{1},f_{1}}^{\dag}a_{\vec{k}_{2},f_{2}}^{\dag}a_{\vec{k}_{3},f_{2}}a_{\vec{k}_{4},f_{1}}\\
&+c_{2}a_{\vec{k}_{1},f_{1}}^{\dag}a_{\vec{k}_{2},f_{2}}^{\dag}\vec{F}_{f_{1}f_{1}^{'}}\cdot\vec{F}_{f_{2}f_{2}^{'}}a_{\vec{k}_{3},f_{2}^{'}}a_{\vec{k}_{4},f_{1}^{'}}),
\end{aligned}
\end{equation}
where $f$ is a hyperfine spin index with value $\pm1,0$, $\vec{F}$ is a standard spin-1 matrix vector, and $\epsilon_{\vec{k}}=\frac{k^{2}}{2m}$ is the kinetic energy of BEC atoms with mass $m$. \\
When an impurity atom also with hyperfine spin 1 is immersed in a spin-1 BEC of a different type atom, the contact potential between the impurity and BEC can be written  \cite{PhysRevLett.114.255301}
\begin{equation}
V_{IB}=\lambda+\beta \vec{F}\cdot\vec{S}+\gamma \hat{P_{0}}.
\end{equation}
Here, $\vec{S}$ represents spin-1 matrix vector of an impurity, $\lambda=(g_{1}+g_{2})/2$ is the spin-independent interaction,
$\beta=(g_{2}-g_{1})/2$ and $\gamma=(2g_{0}-3g_{1}+g_{2})/2$ are the spin-dependent interactions. The strengths of interaction can be expressed as $g_{F}=2\pi a^{IB}_{F}/\mu$ in the first order of perturbation theory, where $a^{IB}_{F}$ is the s-wave scattering length in the total spin $F$ channel with the relative mass $\mu=mM/(m+M)$.  $\hat{P_{0}}$ is a projection operator of spin singlet channels \cite{PhysRevA.86.013632}. Correspondingly, the singlet state of two spin-1 particles is $(\ket{1,-1}-\ket{0,0}+\ket{-1,1})/\sqrt{3}$.

We will focus on the ferromagnetic phase of spin-1 BEC with the order parameter is $(1,0,0)^{T}$, which emerges with the condition $c_{2}<0$ \cite{PhysRevLett.81.742}. It is useful to decompose boson operators as
\begin{equation}
\left(\begin{array}{c}a_{k,1} \\a_{k,0} \\a_{k,-1}\end{array}\right)=\left(\begin{array}{c}\sqrt{n_{0}}+\tilde{a}_{k,1} \\\tilde{a}_{k,0} \\\tilde{a}_{k,-1}\end{array}\right),
\end{equation}
with Bogoliubov transformation \cite{Bogolyubov:1947zz,KAWAGUCHI2012253}
\bea
&&b_{\vec{k},1}=\sqrt{\frac{\epsilon_{\vec{k}}+c_{2}n_{0}+\omega_{\vec{k},1}}{2\omega_{\vec{k},1}}}\tilde{a}_{\vec{k},1}+\sqrt{\frac{\epsilon_{\vec{k}}+c_{2}n_{0}-\omega_{\vec{k},1}}{2\omega_{\vec{k},1}}}\tilde{a}_{-\vec{k},1}^{\dag},\cr&&
b_{\vec{k},0}=\tilde{a}_{\vec{k},0},\cr &&b_{\vec{k},-1}=\tilde{a}_{\vec{k},-1}.\eea
The operators $b_{\vec{k},f}$ are the annihilation operators of three Bogoliubov excitations in spin-1 BEC  with dispersions as the following \cite{PhysRevLett.81.742}: (i) gapless density mode $\omega_{\vec{k},1}=\sqrt{\epsilon_{\vec{k}}(\epsilon_{\vec{k}}+2(c_{0}+c_{2})n_{0})}$,  (ii) ferromagnetic spin wave mode $\omega_{\vec{k},0}=\epsilon_{\vec{k}}$, (iii)
gapped spin quadrupole mode $\omega_{\vec{k},-1}=\epsilon_{\vec{k}}+2|c_{2}|n_{0}$, where  $c_{0}=4\pi (2a^{BB}_{2}+a^{BB}_{0})/(3m)$ and $c_{2}=4\pi(a^{BB}_{2}-a^{BB}_{0})/(3m)$. $a^{BB}_{F}$ is the s-wave scattering length in the total spin $F$ channel.  

Then we can express the total Hamiltonian as
\begin{equation}
H_{tot}=E_{g}+E_{IB}+H_{pol}.
\end{equation}
Here, the first term is the ground state energy of BEC. The second term is the mean-field interaction energy between an impurity and the BEC, which is diagonal in the hyperfine spin states of the impurity as
\begin{equation}
E_{IB}=n_{0}\left(\begin{array}{ccc}\lambda+\beta & 0 & 0 \\0 & \lambda & 0 \\0 & 0 & \lambda-\beta+\frac{\gamma}{3}\end{array}\right)
\end{equation}
The last term has the similar form as Fr\"{o}hlich polaron Hamiltonian\cite{doi:10.1080/00018735400101213}
\begin{equation}
H_{pol}=\frac{p^{2}}{2M}+\sum_{\vec{k},f}\omega_{\vec{k},f}b_{\vec{k},f}^{\dag}b_{\vec{k},f}+\sum_{\vec{k},f}(V_{\vec{k},f}b_{\vec{k},f}e^{i\vec{k}\cdot\vec{r}}+h.c.).\label{Hpol}
\end{equation}
Here and below, we take unit volume and $\hbar=1$. $M$ and $\vec{r}$ are the mass and coordinate of the impurity atom. The ground state of the spinor BEC is treated as a static mean field, above which excitations are modeled as a bath of three different types of boson quasiparticles.
A direct calculation gives coupling matrixes between an impurity atom and collective modes of spinor BEC as
\begin{equation}
\begin{aligned}
V_{\vec{k},1}&=\sqrt{n_{0}W_{k}}\left(
  \begin{array}{ccc}
    \lambda+\beta & 0 & 0\\
    0 & \lambda & 0\\
    0 & 0 & \lambda-\beta+\frac{\gamma}{3}\\
  \end{array}
\right), \\
V_{\vec{k},0}&=\sqrt{n_{0}}\left(
  \begin{array}{ccc}
    0 & 0 & 0\\
    \beta & 0 & 0\\
   0 & \beta-\frac{\gamma}{3} & 0\\
  \end{array}
\right), \\
V_{\vec{k},-1}&=\sqrt{n_{0}}\left(
  \begin{array}{ccc}
    0 & 0 & 0\\
    0 & 0 & 0\\
   \frac{\gamma}{3} &  0 & 0\\
  \end{array}
\right), \label{mat}
\end{aligned}
\end{equation}
where $W_{k}=\frac{\xi k}{\sqrt{2+(\xi k)^{2}}}$ and the healing length of the spinor BEC is defined as $\xi=\frac{1}{\sqrt{2mn_{0}(c_{0}+c_{2})}}$. These matrixes are generalizations of the coupling function in the context of an impurity immersed in a scalar BEC  \cite{:80302,PhysRevB.80.184504}. Correspondingly, relevant collision processes are categorized by the interspecies spin-exchange interactions. There are clear physical meanings for these coupling matrixes.  Since the majority of atoms are condensed in $f=1$ component, the matrix $V_{\vec{k},1}$ has to be diagonal due to the coupling between density wave and an impurity atom.  For spin-exchange processes, we should notice that an impurity and BEC atoms are distinguishable. Without the statistical constrain of  identical bosons, there are two kinds spin-exchange processes corresponding to $V_{\vec{k},0}$ and $V_{\vec{k},-1}$, in contrast to the identical spin-1 case where only one spin-changing process is allowed \cite{PhysRevLett.81.742}. More specifically, $V_{\vec{k},0}$ term exchanges $\hbar$ spin angular momentum accompanied by spin fluctuations, and $V_{\vec{k},-1}$ term exchanges $2\hbar$ spin angular momentum accompanied by spin quadrupolar fluctuations.

We define the dimensionless coupling strengths between an impurity and BEC as
\begin{equation}
\alpha_{f}=\frac{a^{2}_{f}}{a^{BB}_{2}\xi},\label{alp}
\end{equation}
with $a_{1}=a^{IB}_{2}$, $a_{0}=(a^{IB}_{1}+a^{IB}_{2})/2$ and $a_{-1}=(2a^{IB}_{0}+3a^{IB}_{1}+a^{IB}_{2})/6$. Correspondingly, the mean-field energy can be expressed in terms of scattering lengths $a_f$ as $E_{IB,f}=2\pi n_{0}a_{f}/\mu$. It is well known that the Fr\"{o}hlich-like Hamiltonian \eqref{Hpol} resists exact analytical solutions has qualitatively different characteristics in weak-coupling ($\alpha_{f}\ll1$) and strong-coupling ($\alpha_{f}\gg1$) regimes, which will becomes clear in the following investigation.
\section{Perturbative treatments in the weak-coupling regime}\label{weak}
\begin{figure}
\subfigure{\label{2nd}}{
\begin{minipage}{0.48\linewidth}
  \centerline{\includegraphics[width=1.6cm]{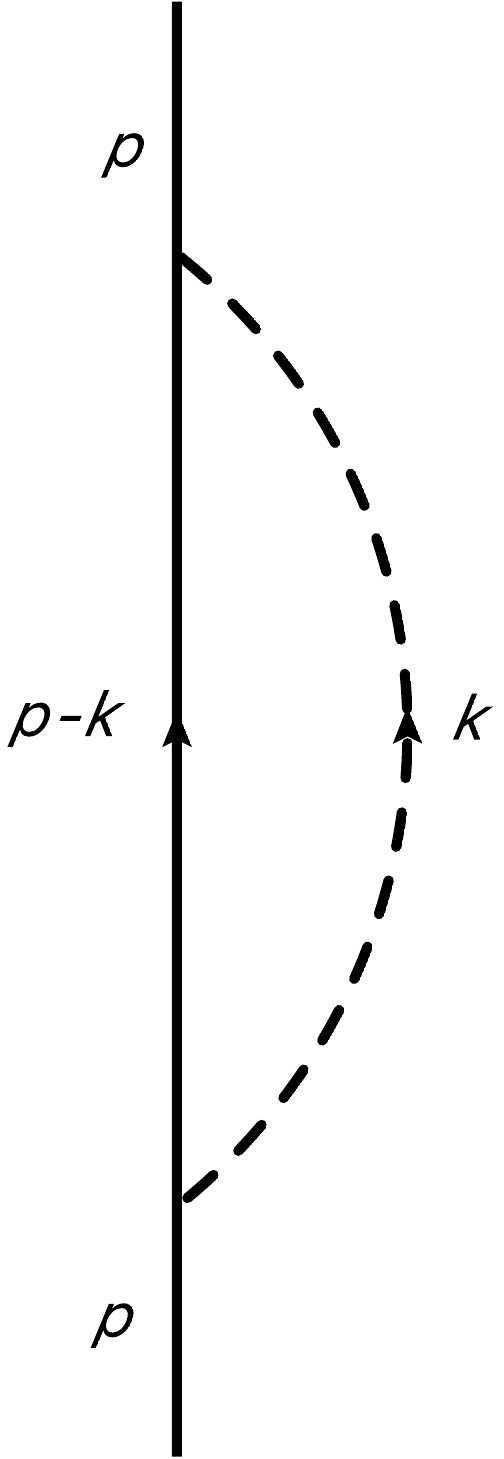}}
  \centerline{(a)}
\end{minipage}}
\hfill
\subfigure{\label{emi}}{
\begin{minipage}{0.48\linewidth}
  \centerline{\includegraphics[width=2.0cm]{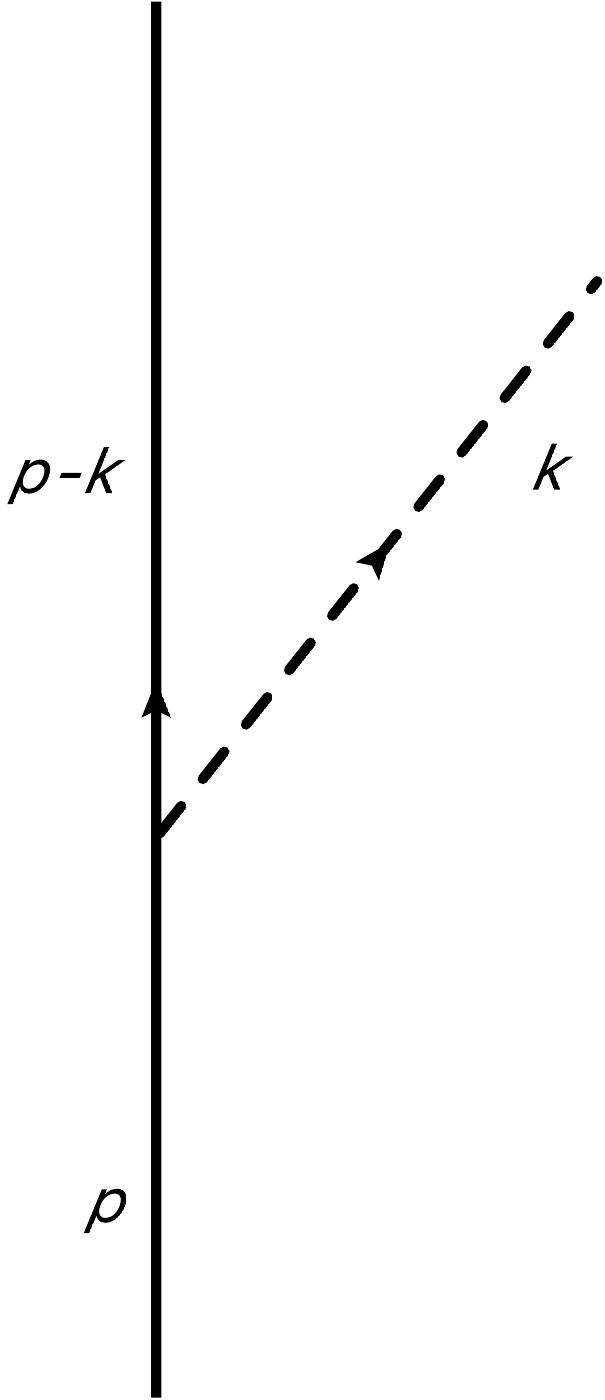}}
  \centerline{(b)}
\end{minipage}}
\caption{(a) The impurity with momentum $\vec{p}$ and virtual collective modes with momentum $k$ corresponds to second-order perturbation theory, where there are three different vertexes coupling to collective modes given by $V_{\vec{k},f}$. (b) The impurity dissipative energy by creating a spin wave mode, which is the only emission process at small momentum. }
\label{fey}
\end{figure}

Let us treat the last term of \eqref{Hpol}
\begin{equation}
H^{\prime}=\sum_{\vec{k},f}(V_{\vec{k},f}b_{\vec{k},f}e^{i\vec{k}\cdot\vec{r}}+h.c.)
\end{equation}
as a perturbation \cite{doi:10.1080/14786445008521794}, which changes the number of quasiparticles. Therefore, the first order perturbed energy vanishes in the vacuum state of quasiparticles.  Considering the second order perturbation processes of different spin components, all initial states have an impurity of momentum $\vec{p}$ and no collective excitations \cite{feynman2018statistical}. In the intermediate state $\ket{n}$, momentum of the impurity and collective excitations are $(\vec{p}-\vec{k})$ and $\vec{k}$, corresponding to the initial energy $E_{0}=\frac{p^{2}}{2M}$ and the intermediate energy $E_{n,f}=\frac{(\vec{p}-\vec{k})^{2}}{2M}+\omega_{\vec{k},f}$, as shown in Fig. \ref{fey}\subref{2nd}. So we obtain the second order perturbed energy
\begin{equation}
\triangle E_{f}=\sum_{n}\frac{|\bra{n}H^{\prime}\ket{0}|^{2}}{E_{0}-E_{n,f}}.
\end{equation}

The effective mass of the impurity can be obtained by performing the small momentum expansion of above equation
\begin{equation}
E_{f}(p)=\triangle E_{f}(0)+\frac{p^{2}}{2M_{f}^{\ast}}+\order{p^{4}}.
\end{equation}
To extract the polaron energy $\triangle E_{f}(0)$ experimentally \cite{PhysRevLett.117.055301,PhysRevLett.117.055302}, we need to take into account the initial mean-field shift $E_{IB}$, because what we actually measure in experiments is  the total energy shift of an impurity due to interactions with BEC as
\begin{equation}
\triangle_{f}=E_{IB,f}+\triangle E_{f}(0),
\end{equation}
For weak interactions, the energies are well described by the mean-field shift. The perturbed corrections can be viewed as the first step beyond the mean-field results. 
For an impurity atom at $f=1$ state, we have
\begin{equation}
\triangle E_{1}(p)=(\lambda+\beta)^{2}n_{0}\sum_{\vec{k}}\frac{W_{k}}{\frac{p^{2}}{2M}-\omega_{\vec{k},1}-\frac{(\vec{p}-\vec{k})^{2}}{2M}}\label{E1},
\end{equation}
which is analogous to the acoustic polaron in solid-state physics.  In the second-order perturbation theory, we must take account of the fact that $g_{2}=\frac{2\pi a^{IB}_{2}}{\mu}$ is not exact, but only valid up to first-order. In the second-order, we have\cite{PhysRevA.89.053617}
\begin{equation}
\frac{2\pi a^{IB}_{2}}{\mu}=g_{2}-g_{2}^{2}\sum_{\vec{k}}\frac{2\mu}{k^{2}}.
\end{equation}
With this cautionary remark, we are able to subtract ultraviolet divergence in  \eqref{E1} from the mean-field shift at the same order, and obtain the finite result of polaron energy at $f=1$ state (Here and below, we take the energy unit as $\frac{1}{M\xi^{2}}$)
 \begin{equation}
\triangle E_{1}(0)=\alpha_{1} I_{1}(\tilde{m}),
\end{equation}
where
\begin{equation}
\begin{aligned}
I_{1}(\tilde{m})&=\frac{(1+\tilde{m})^{2}}{\tilde{m}}\int_{0}^{\infty}\frac{dx}{2\pi}\\
&\times(\frac{1}{1+\tilde{m}}-\frac{x^{2}}{\tilde{m}x\sqrt{2+x^{2}}+2+x^{2}})
\end{aligned}
\end{equation}
with mass ratio $\tilde{m}=m/M$. This is essentially the same results as the polaron in scalar BEC  \cite{PhysRevB.80.184504}. 

For an impurity atom at $f=0$ state, there are two possible intermediates states in the second order perturbation theory as
\begin{equation}
\begin{aligned}
\triangle E_{0}(p) &= \lambda^{2}n_{0}\sum_{\vec{k}}\frac{W_{k}}{\frac{p^{2}}{2M}-\omega_{\vec{k},1}-\frac{(\vec{p}-\vec{k})^{2}}{2M}}\\
&+\beta^{2}n_{0}\sum_{\vec{k}}\frac{1}{\frac{p^{2}}{2M}-\omega_{\vec{k},0}-\frac{(\vec{p}-\vec{k})^{2}}{2M}}.
\end{aligned}
\end{equation}
With the same procedure of subtractions, the energy of a static polaron at $f=0$ state becomes
\begin{equation}
\triangle E_{0}(0)=\alpha_{0} I_{1}(\tilde{m}).
\end{equation}
In contrast to $f=1$ state, no matter how small the initial velocity of the impurity at $f=0$ state is, it dissipates energy by creating a a spin wave mode (magnon) with gapless quadratic dispersion as shown in Fig. \ref{fey}\subref{emi}. In other words, the critical velocity of the impurity at $f=0$ state is zero, which is distinct from the sound velocity at $f=1$ state. In the ferromagnetic phase of spinor BEC, the superfluid current will decay through development of spin textures, due to the spin-gauge symmetry \cite{PhysRevLett.81.742}.

\begin{figure}
    \centering
    \includegraphics[width=8cm]{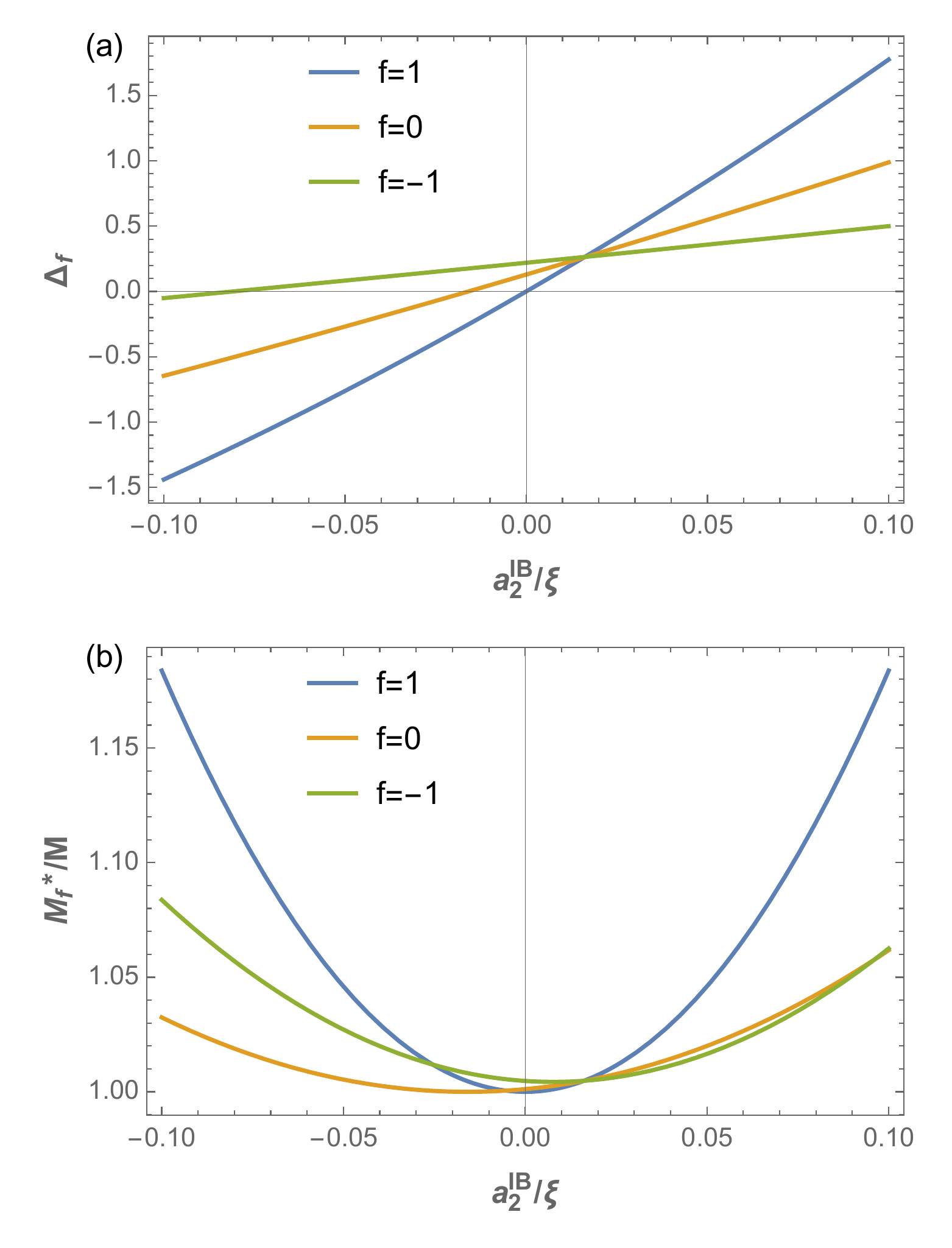}
 \caption{(Color online) Tuning $^{23}$Na-$^{87}$Rb ($\tilde{m}=3.78$) scattering length $a^{IB}_{2}$ while keeping $a^{IB}_{0}=84.1a_{B}$ and $a^{IB}_{1}=81.4a_{B}$ fixed \cite{PhysRevLett.114.255301}, where $a_{B}$ is the Bohr radius. The scattering lengths of $^{87}$Rb are $a^{BB}_{0}=101.8a_{B}$ and $a^{BB}_{2}=100.4a_{B}$ \cite{PhysRevLett.88.093201}. With the typical density of BEC $n_{0}=10^{14}$ cm$^{-3}$, the healing length $\xi=270$ nm. Three coupling constants $\alpha_{f}<0.5$ for the plotting range $|a^{IB}_{2}/\xi|<0.1$, where we expect perturbation theory to be reliable. (a). The total energy shifts of an impurity among different spin components.  (b). Effective mass among different spin components.}
\label{FR}
\end{figure}

There is an additional contribution to polaron energy from the coupling between an impurity and  gapped spin quadrupole excitations for $f=-1$ state, which is analogous to optical polaron in solid-state physics. The second order perturbation theory leads to
\begin{equation}
\begin{aligned}
\triangle E_{-1}(p)&=(\lambda-\beta+\gamma/3)^{2}n_{0}
\sum_{\vec{k}}\frac{W_{k}}{\frac{p^{2}}{2M}-\omega_{\vec{k},1}-\frac{(\vec{p}-\vec{k})^{2}}{2M}}\\
&+(\beta-\frac{\gamma}{3})^{2}n_{0}\sum_{\vec{k}}\frac{1}{\frac{p^{2}}{2M}-\omega_{\vec{k},0}-\frac{(\vec{p}-\vec{k})^{2}}{2M}}\\
&+(\frac{\gamma}{3})^{2}n_{0}\sum_{\vec{k}}\frac{1}{\frac{p^{2}}{2M}-\omega_{\vec{k},-1}-\frac{(\vec{p}-\vec{k})^{2}}{2M}}.\\
\end{aligned}
\end{equation}
The polaron energy at $f=-1$ state can be expressed in terms of a variety of scattering lengths
\begin{equation}
\begin{aligned}
&\triangle E_{-1}(0)=\alpha_{-1} I_{1}(\tilde{m})+\sqrt{\frac{2(a^{BB}_{0}-a^{BB}_{2})}{3a^{BB}_{2}}}\\
&\times(\frac{2a^{IB}_{0}-3a^{IB}_{1}+a^{IB}_{2}}{12a^{IB}_{2}})^{2}\alpha_{1}\frac{\sqrt{1+\tilde{m}}}{\tilde{m}}
\end{aligned}
\end{equation}
Combining the polaron energies with corresponding mean-filed shifts for three spin components, we obtain the total energies of the impurity, which are illustrated in Fig. \ref{FR}(a). We should notice that total energy shifts are dominated by the mean-field contributions in the weak-coupling regime.

Let us now to calculate the effective mass of the impurity for three internal states, which are given by
\begin{equation}
\frac{M_{f}^{\ast}}{M}=\frac{1}{1-M |E_{f}^{''}(0)|}.\label{Meff}
\end{equation}
The singularity at $|E_{f}^{''}(0)|=1/M$ may indicate a disrupt change of ground state in the strong coupling regime, although the perturbation theory has been broken down before the singularity point \cite{MITRA198791}.  The mean-field shift does not contribute to the effective masses, and there are not any ultraviolet divergences in the following calculations.

For an impurity atom at $f=1$ state, a straightforward calculation up to the second order perturbation shows
\begin{equation}
\frac{M_{1}^{\ast}}{M}=1+\alpha_{1} I_{2}(\tilde{m}),
\end{equation}
where
\begin{equation}
\begin{aligned}
I_{2}(\tilde{m})&=\frac{4(1+\tilde{m}^{-1})^{2}}{3\pi}\\
&\times\int_{0}^{\infty}\frac{x^{2}dx}{(\tilde{m}^{-1}\sqrt{2+x^{2}}+x)^{3}\sqrt{2+x^{2}}}.
\end{aligned}
\end{equation}
The enhancement of effective mass relative to the bare mass  indicates the lower mobility of the impurity, due to the formation of screening quasiparticles around. Correspondingly, the effective mass of an impurity atom at $f=0$ and $f=-1$ states are
\begin{equation}
\begin{aligned}
&\frac{M_{0}^{\ast}}{M}=1+\alpha_{0} I_{2}(\tilde{m}),\\
&\frac{M_{-1}^{\ast}}{M}=1+\alpha_{-1} I_{2}(\tilde{m})+\sqrt{\frac{3a^{BB}_{2}}{2(a^{BB}_{0}-a^{BB}_{2})}}\\
&\times(\frac{2a^{IB}_{0}-3a^{IB}_{1}+a^{IB}_{2}}{12a^{IB}_{2}})^{2} \alpha_{1}\frac{\tilde{m}}{\sqrt{1+\tilde{m}}}.
\end{aligned}
\end{equation}
In the weak-coupling regime, the splittings of effective mass among different spin components are within a few percent as shown in Fig. \ref{FR}(b). We should notice that these results are coincident with the intermediate coupling theory of Lee-Low-Pines up to the order of $\alpha_{f}$ \cite{PhysRev.90.297}.  Mobility of the impurity depends on the hyperfine spin states, since the screening effects by the surrounding collective modes depends on the different types of dispersion.  The splitting effective mass of different spin states is a characteristic phenomena in spin-polaron physics, which plays an important role in strongly correlated electron materials \cite{PhysRev.118.141,Alexandrov2007PolaronsIA}.

\section{Polaron energy and effective mass in the strong-coupling regime}\label{strong}
Beginning with Landau's speculation that an electron in an ionic crystal can be trapped by lattice distortion produced by the electron itself \cite{Landau:1933iwn}, subsequent works gave rise to the strong-coupling polaron theory. Following the approach of Pekar and Landau \cite{Pekar}, we suggest a product ansatz for the trial wave function of Hamiltonian  \eqref{Hpol}
\begin{equation}
\ket{\Psi}=\ket{\phi}\ket{\chi}\ket{\psi_{1}}\ket{\psi_{0}}\ket{\psi_{-1}},
\end{equation}
where $\ket{\phi}$ and $\ket{\chi}$ are normalized spatial and spinor functions of the impurity variable only, and $\ket{\psi_{f}}$ that of collective modes variables only. The variational principle requires the minimization of energy
\begin{equation}
\begin{aligned}
\expval{H_{pol}}{\Psi}=&-\expval{\frac{\nabla^{2}}{2M}}{\phi}+\sum_{\vec{k},f}\omega_{f}(\vec{k})\expval{b^{\dag}_{\vec{k}f}b_{\vec{k}f}}{\psi_{f}}\\
&+\sum_{\vec{k},f}(G^{\ast}_{f}(\vec{k})\expval{b_{\vec{k}f}}{\psi_{f}}+h.c.),
\end{aligned}
\end{equation}
with the definition
\begin{equation}
G^{\ast}_{f}(\vec{k})\equiv \expval{e^{i\vec{k}\cdot\vec{r}}}{\phi}\expval{V_{\vec{k}f}}{\chi}.
\end{equation}
It is convenient to introduce displaced collective modes operators
\begin{equation}
B_{f}(\vec{k})\equiv b_{\vec{k}f}+\frac{G_{f}(\vec{k})}{\omega_{f}(\vec{k})},
\end{equation}
which satisfies the canonical commutation rules. Then the polaron energy is calculated as
\begin{equation}
\begin{aligned}
\expval{H_{pol}}{\Psi}=&-\expval{\frac{\nabla^{2}}{2M}}{\phi}-\sum_{\vec{k},f}\frac{|G_{f}(\vec{k})|^{2}}{\omega_{f}(\vec{k})}\\
&+\sum_{\vec{k},f}\omega_{f}(\vec{k})\expval{B^{\dag}_{\vec{k}f}B_{\vec{k}f}}{\psi_{f}}.
\end{aligned}
\end{equation}
The second term is always nonnegative, because $B^{\dag}_{\vec{k}f}B_{\vec{k}f}$ is a quadratic Hermitian operator and $\omega_{f}(\vec{k})\geqslant0$ is required by the thermodynamic stability at absolute zero \cite{Bogolyubov:1947zz}. Choosing the variational state function of collective modes $\ket{\psi^{0}_{f}}$ satisfying $B_{f}(\vec{k})\ket{\psi^{0}_{f}}=0$, the variational energy becomes
\begin{equation}
\expval{H_{pol}}{\Psi}=-\expval{\frac{\nabla^{2}}{2M}}{\phi}-\sum_{\vec{k},f}\frac{|G_{f}(\vec{k})|^{2}}{\omega_{f}(\vec{k})}.\label{vpol}
\end{equation}
Generally, a spinor wave function $\ket{\chi}$ of the impurity is a superposition of three internal states. In the following, we only concern about the impurity at a hyperfine spin eigenstate, where the last term of \eqref{vpol}  is determined by the diagonal elements of $V_{\vec{k}1}$.  In order to evaluate the ground state energy, we have to specify the concrete form of trial wave function $\phi(\vec{r})$  of the impurity. The simplest choice is a hydrogen-like ground state wave function \cite{doi:10.1080/00018735400101213}
\begin{equation}
\phi(\vec{r})=\sqrt{\frac{t^{3}}{\pi\xi^{3}}}e^{-tr/\xi},\label{wave}
\end{equation}
where $t$ is a dimensionless variational parameter related to the radius of polaron
\begin{equation}
\frac{\sqrt{<r^{2}>}}{\xi}=\frac{\sqrt{3}}{t}.\label{rad}
\end{equation}
Substituting \eqref{wave} into \eqref{vpol}, the variational energy for an impurity at a spin $f$ eigenstate can be evaluated straightforwardly as
\begin{equation}
\mathcal{E}_{f}=\frac{t^{2}}{2}-\tilde{\alpha}_{f}I_{3}(t),\label{varE}
\end{equation}
where $\tilde{\alpha}_{f}\equiv\alpha_{f} (1+\tilde{m})(1+\tilde{m}^{-1})$ and
\begin{equation}
I_{3}(t)=\frac{2}{\pi}t^{3}\int_{0}^{\infty}\frac{x^{2}dx}{(1+x^{2})^{4}(1+2t^{2}x^{2})}.
\end{equation}

\begin{figure}
    \centering
    \includegraphics[width=8cm]{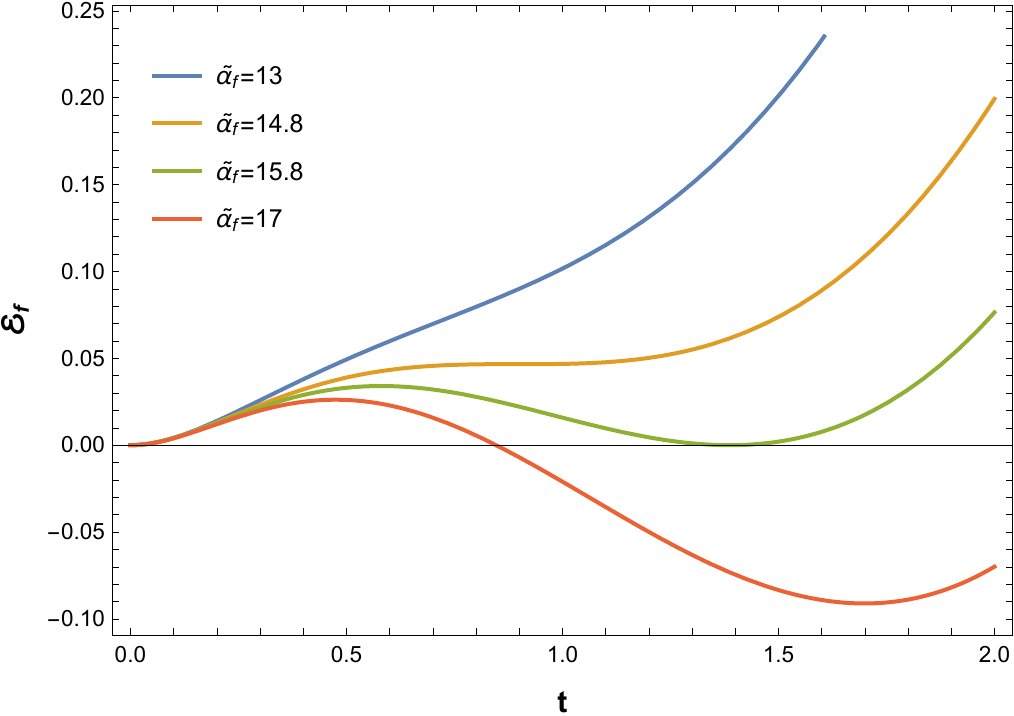}
\caption{(Color online) The variational energy is shown as a function of the variational parameter at different coupling strength. The variational energy start to develop a local minimum when $\tilde{\alpha}_{f}>14.8$. The polaron energy becomes negative at a critical coupling strength $\tilde{\alpha}_{f}=15.8$, which indicates a transition from an unbound state to a localized state. Corresponding to the minimum of energy, there is a discontinuity on the variational parameter $t$, jumping from zero to 1.39.}
\label{conE}
\end{figure}

\begin{figure}
    \centering
    \includegraphics[width=8cm]{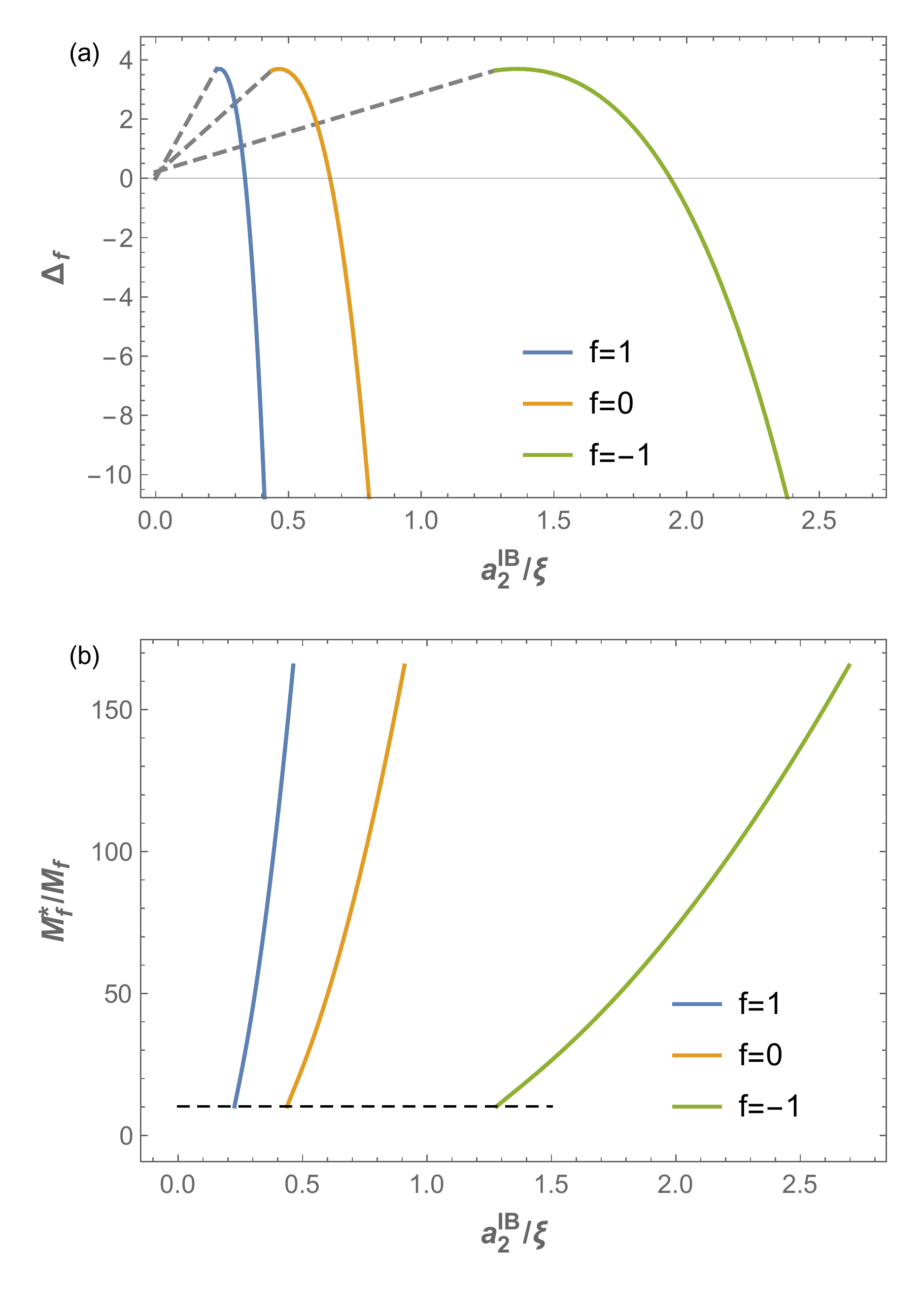}
 \caption{(Color online) Relevant parameters are chosen as Fig.\ref{FR}. (a). The energy shifts of an impurity among different spin components are plotted as functions of the scattering length in the strong-coupling regime. Dashed lines indicate the mean-field shift before critical coupling strengths.  (b). Effective mass among different spin components. The horizontal dashed line at $M_{f}^{\ast}/M=10.2$ indicates discontinuities at critical coupling strengths. In the strong-coupling regime, the splitting of effective mass among different spin components are appreciable.}
\label{STR}
\end{figure}

The strong-coupling polaron corresponds to the solutions of negative variational energy, which is determined by the value of coupling constant $\tilde{\alpha}_{f}$. Only when $\tilde{\alpha}_{f}$ exceeds some critical value, the minimum of variational energy will be lower than zero. The impurity atom can transit from a mobile state to a localized state, so call self-trapping state. This essential feature is shown in Fig.\ref{conE}, where the variational treatment predicts that the impurity start to localize at the critical coupling strength
\begin{equation}
\alpha_{f}^{\ast}=\frac{15.8}{(1+\tilde{m})(1+\tilde{m}^{-1})}.\label{cr}
\end{equation}
Let us consider a much lower density gas of spinor impurity atoms immersed in a spinor BEC.  The impurity -BEC interactions are characterized by three scattering lengths, while the interactions among impurity atoms are safely neglected. Inferring from \eqref{cr} and \eqref{alp}, we can manipulate the formation of polarons for different spin components by tuning the heteronuclear scattering lengths. As shown in Fig. \ref{STR}(a) for a $^{23}$Na-$^{87}$Rb system, we plot the total energy of the impurity defined as $\triangle_{f}=E_{IB,f}+\mathcal{E}_{f}$. Increasing $a^{IB}_{2}$  while keeping $a^{IB}_{0}$ and $a^{IB}_{1}$ fixed,  the polaron of $f=1$ will first develop a localized state at $a^{IB}_{2}/\xi=0.23$. Polarons later becomes localized at  $a^{IB}_{2}/\xi=0.44$ and $a^{IB}_{2}/\xi=1.28$ for $f=0$ and $f=-1$, respectively.  From \eqref{rad}, polarons can self-localize in a region smaller than the healing length, which could also be identified as the system crossing over from the weak to strong coupling regime. 

The above solution of the polaron energy is static, where the impurity is centered around origin and trapped by BEC density distortion. In order to study the effective mass in strong coupling regime, it is assumed that the impurity state function $\ket{\tilde{\phi}}$ moves with a small velocity $\vec{u}$. Then we minimize the following quantity \cite{DAPROVIDENCIA1975366}
\begin{equation}
\begin{aligned}
&\expval{H_{pol}-\vec{u}\cdot\vec{P}}{\tilde{\Psi}}=-\expval{\frac{\nabla^{2}}{2M}}{\tilde{\phi}}+i\vec{u}\cdot\expval{\nabla}{\tilde{\phi}}\\
&+\sum_{\vec{k},f}(G^{\ast}_{f}(\vec{k})\expval{b_{\vec{k}f}}{\psi_{f}}+h.c.)\\
&+\sum_{\vec{k},f}(\omega_{f}(\vec{k})-\vec{u}\cdot\vec{k})\expval{b^{\dag}_{\vec{k}f}b_{\vec{k}f}}{\psi_{f}},
\end{aligned}
\end{equation}
where the Lagrange multiplier $\vec{u}$ is the velocity of polaron and the total momentum operator is $\vec{P}=-i\hbar\nabla+\sum_{\vec{k},f}\vec{k}b_{\vec{k},f}^{\dag}b_{\vec{k},f}$. Redefining the displaced operator $\tilde{B}_{f}(\vec{k})\equiv b_{\vec{k}f}+\frac{G_{f}(\vec{k})}{\omega_{f}(\vec{k})-\vec{u}\cdot\vec{k}}$,
the phonon ground state is solved by $\tilde{B}_{f}(\vec{k})\ket{\tilde{\psi}^{0}_{f}}=0$. Then, the variational energy up to the second order of velocity is found to be
\begin{equation}
\begin{aligned}
&\mathcal{E}_{f}(\vec{u})=-\expval{\frac{\nabla^{2}}{2M}}{\phi}+\frac{1}{2}Mu^{2}\\
&-\sum_{\vec{k},f}\frac{|G_{f}(\vec{k})|^{2}}{\omega_{f}(\vec{k})}+\sum_{\vec{k},f}(\vec{u}\cdot\vec{k})^{2}\frac{|G_{f}(\vec{k})|^{2}}{\omega^{3}_{f}(\vec{k})}.
\end{aligned}\label{varEu}
\end{equation}
The effective mass of spin $f$ eigenstate can be expressed in terms of the variational parameter as
\begin{equation}
\frac{M_{f}^{\ast}}{M}=1+\tilde{m}^{2}\tilde{\alpha}_{f}I_{4}(t),\label{effM}
\end{equation}
where
\begin{equation}
I_{4}(t)=\frac{8}{3\pi}t^{3}\int_{0}^{\infty}\frac{x^{2}dx}{(1+x^{2})^{4}(1+2t^{2}x^{2})^{2}}.
\end{equation}
By the minimization of variation energy \eqref{varE}, we could obtain the best choice of variational parameter $t$ for every coupling constant $\tilde{\alpha}_{f}$ above the critical value. As shown in  Fig. \ref{STR}(b) , the polarons effective mass can increase with several order of magnitudes, in contrast with the weak-coupling regime where the enhancements of effective mass are within few percents.  The large splittings of effective mass among different spin components are manifested on the strong-coupling regime, which could be detected by measuring the mobility of impurity. We should also notice that there is a stronger enhancement of the effective mass for a lighter impurity, due to the factor $\tilde{m}^{2}$ in \eqref{effM}.  It has been pointed out that the large effective mass regime is difficult to realize in most solid materials \cite{PhysRevB.32.3515}.  Therefore, the large scale enhancement of the effective mass in the ultracold gases will open up the prospect to investigate this regime hitherto inaccessible in the solids.

\section{Discussions and summaries}\label{exp}
In the current experimental conditions, we estimate typical parameters for our system as following. The most important one is the dimensionless coupling constant $\alpha_{f}$ defined by \eqref{alp}. For the s-wave scattering  between spinor gases of $^{23}$Na and $^{87}$Rb, it has been measured that $(\lambda,\beta,\gamma)=2\pi\hbar^{2}a_{B}/\mu\times(78.9,-2.5,0.06)$ \cite{PhysRevLett.114.255301}, where $a_{B}$ is the Bohr radius. With the typical density of BEC $n_{0}=10^{14}$ cm$^{-3}$ and scattering lengths  $a^{BB}_{2}=100.4a_{B}$ \cite{PhysRevLett.88.093201}, we notice that the healing length $\xi\approx270$ nm and the energy unit of polaron $1/(M\xi^{2})\approx290$ nK. For these values, we obtain the coupling constants $\alpha_{f}\approx0.01$, so the weak-coupling perturbation theory is suitable.  Since both $\beta$ and $\gamma$ are much smaller than $\lambda$ in this experimental setup, the splitting of polaron energies and effective mass among different hyperfine spin states are within few percents. 

We have assumed spatial homogeneity in our investigation so far. Let us consider the effect of a harmonic trap on the dimensionless coupling constant $\alpha_{f}$. The density of BEC is well approximated by the Thomas-Fermi approximation for a large cloud \cite{pethick_smith_2008}. It is straightforward to show that the radius of cloud $R\propto (a^{BB}_{2}/\bar{a})^{1/5}\bar{a}$, where $\bar{a}=1/\sqrt{M\bar{\omega}}$ is a trap length and $\bar{\omega}$ is a trap frequency. Therefore, we find the coupling constant $\alpha_{f}$ in the presence of a harmonic trap as $\alpha_{f}\propto(a^{BB}_{2}/\bar{a})^{1/5}(a_{f}^{2}/a^{BB}_{2}\bar{a})$. In order to explore the properties of polarons in strong-coupling regime, there are three ways to enhance the coupling strength, which are increasing impurity-BEC scattering length $a_{f}$, decreasing BEC scattering length $a^{BB}_{2}$, or tighten up the harmonic trap \cite{PhysRevB.80.184504}. Experimentally, all the three ways can be well explored. In our proposals, we mainly focus on tuning $a^{IB}_{2}$ with current techniques of the optical Feshbach resonance for spinor gases.  For example, when we are able to increase $a^{IB}_{2}$ up to a factor of 10, we enhance the effective coupling constant $\alpha_{f}$ up to a factor of about 100, which could provide an experimental access to the strong-coupling regime $\alpha_{f}\approx1$.   We should notice that it is much more challenging to control the spinor system near the resonance. An alternative technique to tune the spin-dependent interaction is to use the microwave-induced Feshbach resonance \cite{PhysRevA.81.041603}, in order to reduce atomic losses near the resonance. Although there is still no such kind of measurement performed in a spinor impurity-BEC system, we expect that the large splittings of polaron energies and effective masses among different spin components in the strong-coupling regime will be observed in the near future experiment. 

In summary, we reveal polaron effects where a spin-1 impurity atom immersed in a ferromagnetic phase of spinor BEC. In the weak-coupling regime, the contributions from different types of collective modes are studied with the perturbation theory. In the strong-coupling regime, the gapless phonon mode plays a dominant role in the formation of self-trapping states. We predict critical values of coupling constants by a variational treatment, where the impurity start to transit from a mobile state to a localized state. The properties of polarons are best manifested on the splittings of energies and effective masses among different spin components. Our results are of particular significance for exploring the new features of polaron effects in spinor BEC, which are very distinct from the scalar case.

\section{Acknowledgments}
The work is supported by the Beijing Natural Science Foundation (Grand No. Z180007).
\bibliography{polaron1.bib}

\begin{thebibliography}{46}%
\makeatletter
\providecommand \@ifxundefined [1]{%
 \@ifx{#1\undefined}
}%
\providecommand \@ifnum [1]{%
 \ifnum #1\expandafter \@firstoftwo
 \else \expandafter \@secondoftwo
 \fi
}%
\providecommand \@ifx [1]{%
 \ifx #1\expandafter \@firstoftwo
 \else \expandafter \@secondoftwo
 \fi
}%
\providecommand \natexlab [1]{#1}%
\providecommand \enquote  [1]{``#1''}%
\providecommand \bibnamefont  [1]{#1}%
\providecommand \bibfnamefont [1]{#1}%
\providecommand \citenamefont [1]{#1}%
\providecommand \href@noop [0]{\@secondoftwo}%
\providecommand \href [0]{\begingroup \@sanitize@url \@href}%
\providecommand \@href[1]{\@@startlink{#1}\@@href}%
\providecommand \@@href[1]{\endgroup#1\@@endlink}%
\providecommand \@sanitize@url [0]{\catcode `\\12\catcode `\$12\catcode
  `\&12\catcode `\#12\catcode `\^12\catcode `\_12\catcode `\%12\relax}%
\providecommand \@@startlink[1]{}%
\providecommand \@@endlink[0]{}%
\providecommand \url  [0]{\begingroup\@sanitize@url \@url }%
\providecommand \@url [1]{\endgroup\@href {#1}{\urlprefix }}%
\providecommand \urlprefix  [0]{URL }%
\providecommand \Eprint [0]{\href }%
\providecommand \doibase [0]{https://doi.org/}%
\providecommand \selectlanguage [0]{\@gobble}%
\providecommand \bibinfo  [0]{\@secondoftwo}%
\providecommand \bibfield  [0]{\@secondoftwo}%
\providecommand \translation [1]{[#1]}%
\providecommand \BibitemOpen [0]{}%
\providecommand \bibitemStop [0]{}%
\providecommand \bibitemNoStop [0]{.\EOS\space}%
\providecommand \EOS [0]{\spacefactor3000\relax}%
\providecommand \BibitemShut  [1]{\csname bibitem#1\endcsname}%
\let\auto@bib@innerbib\@empty
\bibitem [{\citenamefont {Fröhlich}(1954)}]{doi:10.1080/00018735400101213}%
  \BibitemOpen
  \bibfield  {author} {\bibinfo {author} {\bibfnamefont {H.}~\bibnamefont
  {Fröhlich}},\ }\href {https://doi.org/10.1080/00018735400101213} {\bibfield
  {journal} {\bibinfo  {journal} {Adv. Phys.}\ }\textbf {\bibinfo {volume}
  {3}},\ \bibinfo {pages} {325} (\bibinfo {year} {1954})}\BibitemShut {NoStop}%
\bibitem [{\citenamefont {Mahan}(1990)}]{mahan1990many}%
  \BibitemOpen
  \bibfield  {author} {\bibinfo {author} {\bibfnamefont {G.}~\bibnamefont
  {Mahan}},\ }\href@noop {} {\emph {\bibinfo {title} {Many-Particle
  Physics}}},\ Physics of Solids and Liquids\ (\bibinfo  {publisher} {Springer
  US},\ \bibinfo {year} {1990})\BibitemShut {NoStop}%
\bibitem [{\citenamefont {Miller}\ \emph {et~al.}(1962)\citenamefont {Miller},
  \citenamefont {Pines},\ and\ \citenamefont {Nozi\`eres}}]{PhysRev.127.1452}%
  \BibitemOpen
  \bibfield  {author} {\bibinfo {author} {\bibfnamefont {A.}~\bibnamefont
  {Miller}}, \bibinfo {author} {\bibfnamefont {D.}~\bibnamefont {Pines}},\ and\
  \bibinfo {author} {\bibfnamefont {P.}~\bibnamefont {Nozi\`eres}},\ }\href
  {https://doi.org/10.1103/PhysRev.127.1452} {\bibfield  {journal} {\bibinfo
  {journal} {Phys. Rev.}\ }\textbf {\bibinfo {volume} {127}},\ \bibinfo {pages}
  {1452} (\bibinfo {year} {1962})}\BibitemShut {NoStop}%
\bibitem [{\citenamefont {Hernandez}(1991)}]{RevModPhys.63.675}%
  \BibitemOpen
  \bibfield  {author} {\bibinfo {author} {\bibfnamefont {J.~P.}\ \bibnamefont
  {Hernandez}},\ }\href {https://doi.org/10.1103/RevModPhys.63.675} {\bibfield
  {journal} {\bibinfo  {journal} {Rev. Mod. Phys.}\ }\textbf {\bibinfo {volume}
  {63}},\ \bibinfo {pages} {675} (\bibinfo {year} {1991})}\BibitemShut
  {NoStop}%
\bibitem [{\citenamefont {Alexandrov}(2007)}]{Alexandrov2007PolaronsIA}%
  \BibitemOpen
  \bibfield  {author} {\bibinfo {author} {\bibfnamefont {A.}~\bibnamefont
  {Alexandrov}},\ }\href@noop {} {\emph {\bibinfo {title} {Polarons in Advanced
  Materials}}}\ (\bibinfo  {publisher} {Springer},\ \bibinfo {year}
  {2007})\BibitemShut {NoStop}%
\bibitem [{\citenamefont {Emin}(2012)}]{emin_2012}%
  \BibitemOpen
  \bibfield  {author} {\bibinfo {author} {\bibfnamefont {D.}~\bibnamefont
  {Emin}},\ }\href {https://doi.org/10.1017/CBO9781139023436} {\emph {\bibinfo
  {title} {Polarons}}}\ (\bibinfo  {publisher} {Cambridge University Press},\
  \bibinfo {year} {2012})\BibitemShut {NoStop}%
\bibitem [{\citenamefont {Bredas}\ and\ \citenamefont
  {Street}(1985)}]{doi:10.1021/ar00118a005}%
  \BibitemOpen
  \bibfield  {author} {\bibinfo {author} {\bibfnamefont {J.~L.}\ \bibnamefont
  {Bredas}}\ and\ \bibinfo {author} {\bibfnamefont {G.~B.}\ \bibnamefont
  {Street}},\ }\href {https://doi.org/10.1021/ar00118a005} {\bibfield
  {journal} {\bibinfo  {journal} {Acc. Chem. Res.}\ }\textbf {\bibinfo {volume}
  {18}},\ \bibinfo {pages} {309} (\bibinfo {year} {1985})}\BibitemShut
  {NoStop}%
\bibitem [{\citenamefont {Scott}(1992)}]{SCOTT19921}%
  \BibitemOpen
  \bibfield  {author} {\bibinfo {author} {\bibfnamefont {A.}~\bibnamefont
  {Scott}},\ }\href
  {https://doi.org/https://doi.org/10.1016/0370-1573(92)90093-F} {\bibfield
  {journal} {\bibinfo  {journal} {Phys. Rep.}\ }\textbf {\bibinfo {volume}
  {217}},\ \bibinfo {pages} {1 } (\bibinfo {year} {1992})}\BibitemShut
  {NoStop}%
\bibitem [{\citenamefont {Conwell}(2005)}]{Conwell8795}%
  \BibitemOpen
  \bibfield  {author} {\bibinfo {author} {\bibfnamefont {E.~M.}\ \bibnamefont
  {Conwell}},\ }\href {https://doi.org/10.1073/pnas.0501406102} {\bibfield
  {journal} {\bibinfo  {journal} {Proc. Natl. Acad. Sci.}\ }\textbf {\bibinfo
  {volume} {102}},\ \bibinfo {pages} {8795} (\bibinfo {year}
  {2005})}\BibitemShut {NoStop}%
\bibitem [{\citenamefont {Hu}\ \emph {et~al.}(2016)\citenamefont {Hu},
  \citenamefont {Van~de Graaff}, \citenamefont {Kedar}, \citenamefont {Corson},
  \citenamefont {Cornell},\ and\ \citenamefont {Jin}}]{PhysRevLett.117.055301}%
  \BibitemOpen
  \bibfield  {author} {\bibinfo {author} {\bibfnamefont {M.-G.}\ \bibnamefont
  {Hu}}, \bibinfo {author} {\bibfnamefont {M.~J.}\ \bibnamefont {Van~de
  Graaff}}, \bibinfo {author} {\bibfnamefont {D.}~\bibnamefont {Kedar}},
  \bibinfo {author} {\bibfnamefont {J.~P.}\ \bibnamefont {Corson}}, \bibinfo
  {author} {\bibfnamefont {E.~A.}\ \bibnamefont {Cornell}},\ and\ \bibinfo
  {author} {\bibfnamefont {D.~S.}\ \bibnamefont {Jin}},\ }\href
  {https://doi.org/10.1103/PhysRevLett.117.055301} {\bibfield  {journal}
  {\bibinfo  {journal} {Phys. Rev. Lett.}\ }\textbf {\bibinfo {volume} {117}},\
  \bibinfo {pages} {055301} (\bibinfo {year} {2016})}\BibitemShut {NoStop}%
\bibitem [{\citenamefont {J\o{}rgensen}\ \emph {et~al.}(2016)\citenamefont
  {J\o{}rgensen}, \citenamefont {Wacker}, \citenamefont {Skalmstang},
  \citenamefont {Parish}, \citenamefont {Levinsen}, \citenamefont
  {Christensen}, \citenamefont {Bruun},\ and\ \citenamefont
  {Arlt}}]{PhysRevLett.117.055302}%
  \BibitemOpen
  \bibfield  {author} {\bibinfo {author} {\bibfnamefont {N.~B.}\ \bibnamefont
  {J\o{}rgensen}}, \bibinfo {author} {\bibfnamefont {L.}~\bibnamefont
  {Wacker}}, \bibinfo {author} {\bibfnamefont {K.~T.}\ \bibnamefont
  {Skalmstang}}, \bibinfo {author} {\bibfnamefont {M.~M.}\ \bibnamefont
  {Parish}}, \bibinfo {author} {\bibfnamefont {J.}~\bibnamefont {Levinsen}},
  \bibinfo {author} {\bibfnamefont {R.~S.}\ \bibnamefont {Christensen}},
  \bibinfo {author} {\bibfnamefont {G.~M.}\ \bibnamefont {Bruun}},\ and\
  \bibinfo {author} {\bibfnamefont {J.~J.}\ \bibnamefont {Arlt}},\ }\href
  {https://doi.org/10.1103/PhysRevLett.117.055302} {\bibfield  {journal}
  {\bibinfo  {journal} {Phys. Rev. Lett.}\ }\textbf {\bibinfo {volume} {117}},\
  \bibinfo {pages} {055302} (\bibinfo {year} {2016})}\BibitemShut {NoStop}%
\bibitem [{\citenamefont {Mathey}\ \emph {et~al.}(2004)\citenamefont {Mathey},
  \citenamefont {Wang}, \citenamefont {Hofstetter}, \citenamefont {Lukin},\
  and\ \citenamefont {Demler}}]{PhysRevLett.93.120404}%
  \BibitemOpen
  \bibfield  {author} {\bibinfo {author} {\bibfnamefont {L.}~\bibnamefont
  {Mathey}}, \bibinfo {author} {\bibfnamefont {D.-W.}\ \bibnamefont {Wang}},
  \bibinfo {author} {\bibfnamefont {W.}~\bibnamefont {Hofstetter}}, \bibinfo
  {author} {\bibfnamefont {M.~D.}\ \bibnamefont {Lukin}},\ and\ \bibinfo
  {author} {\bibfnamefont {E.}~\bibnamefont {Demler}},\ }\href
  {https://doi.org/10.1103/PhysRevLett.93.120404} {\bibfield  {journal}
  {\bibinfo  {journal} {Phys. Rev. Lett.}\ }\textbf {\bibinfo {volume} {93}},\
  \bibinfo {pages} {120404} (\bibinfo {year} {2004})}\BibitemShut {NoStop}%
\bibitem [{\citenamefont {Huang}\ and\ \citenamefont {Wan}(2009)}]{:80302}%
  \BibitemOpen
  \bibfield  {author} {\bibinfo {author} {\bibfnamefont {B.-B.}\ \bibnamefont
  {Huang}}\ and\ \bibinfo {author} {\bibfnamefont {S.-L.}\ \bibnamefont
  {Wan}},\ }\href {https://doi.org/10.1088/0256-307X/26/8/080302} {\bibfield
  {journal} {\bibinfo  {journal} {Chin. Phys. Lett.}\ }\textbf {\bibinfo
  {volume} {26}},\ \bibinfo {eid} {80302} (\bibinfo {year} {2009})}\BibitemShut
  {NoStop}%
\bibitem [{\citenamefont {Tempere}\ \emph {et~al.}(2009)\citenamefont
  {Tempere}, \citenamefont {Casteels}, \citenamefont {Oberthaler},
  \citenamefont {Knoop}, \citenamefont {Timmermans},\ and\ \citenamefont
  {Devreese}}]{PhysRevB.80.184504}%
  \BibitemOpen
  \bibfield  {author} {\bibinfo {author} {\bibfnamefont {J.}~\bibnamefont
  {Tempere}}, \bibinfo {author} {\bibfnamefont {W.}~\bibnamefont {Casteels}},
  \bibinfo {author} {\bibfnamefont {M.~K.}\ \bibnamefont {Oberthaler}},
  \bibinfo {author} {\bibfnamefont {S.}~\bibnamefont {Knoop}}, \bibinfo
  {author} {\bibfnamefont {E.}~\bibnamefont {Timmermans}},\ and\ \bibinfo
  {author} {\bibfnamefont {J.~T.}\ \bibnamefont {Devreese}},\ }\href
  {https://doi.org/10.1103/PhysRevB.80.184504} {\bibfield  {journal} {\bibinfo
  {journal} {Phys. Rev. B}\ }\textbf {\bibinfo {volume} {80}},\ \bibinfo
  {pages} {184504} (\bibinfo {year} {2009})}\BibitemShut {NoStop}%
\bibitem [{\citenamefont {Cucchietti}\ and\ \citenamefont
  {Timmermans}(2006)}]{PhysRevLett.96.210401}%
  \BibitemOpen
  \bibfield  {author} {\bibinfo {author} {\bibfnamefont {F.~M.}\ \bibnamefont
  {Cucchietti}}\ and\ \bibinfo {author} {\bibfnamefont {E.}~\bibnamefont
  {Timmermans}},\ }\href {https://doi.org/10.1103/PhysRevLett.96.210401}
  {\bibfield  {journal} {\bibinfo  {journal} {Phys. Rev. Lett.}\ }\textbf
  {\bibinfo {volume} {96}},\ \bibinfo {pages} {210401} (\bibinfo {year}
  {2006})}\BibitemShut {NoStop}%
\bibitem [{\citenamefont {Yoshida}\ \emph {et~al.}(2018)\citenamefont
  {Yoshida}, \citenamefont {Endo}, \citenamefont {Levinsen},\ and\
  \citenamefont {Parish}}]{PhysRevX.8.011024}%
  \BibitemOpen
  \bibfield  {author} {\bibinfo {author} {\bibfnamefont {S.~M.}\ \bibnamefont
  {Yoshida}}, \bibinfo {author} {\bibfnamefont {S.}~\bibnamefont {Endo}},
  \bibinfo {author} {\bibfnamefont {J.}~\bibnamefont {Levinsen}},\ and\
  \bibinfo {author} {\bibfnamefont {M.~M.}\ \bibnamefont {Parish}},\ }\href
  {https://doi.org/10.1103/PhysRevX.8.011024} {\bibfield  {journal} {\bibinfo
  {journal} {Phys. Rev. X}\ }\textbf {\bibinfo {volume} {8}},\ \bibinfo {pages}
  {011024} (\bibinfo {year} {2018})}\BibitemShut {NoStop}%
\bibitem [{\citenamefont {Wang}\ \emph {et~al.}(2019)\citenamefont {Wang},
  \citenamefont {Liu},\ and\ \citenamefont {Hu}}]{PhysRevLett.123.213401}%
  \BibitemOpen
  \bibfield  {author} {\bibinfo {author} {\bibfnamefont {J.}~\bibnamefont
  {Wang}}, \bibinfo {author} {\bibfnamefont {X.-J.}\ \bibnamefont {Liu}},\ and\
  \bibinfo {author} {\bibfnamefont {H.}~\bibnamefont {Hu}},\ }\href
  {https://doi.org/10.1103/PhysRevLett.123.213401} {\bibfield  {journal}
  {\bibinfo  {journal} {Phys. Rev. Lett.}\ }\textbf {\bibinfo {volume} {123}},\
  \bibinfo {pages} {213401} (\bibinfo {year} {2019})}\BibitemShut {NoStop}%
\bibitem [{\citenamefont {Rath}\ and\ \citenamefont
  {Schmidt}(2013)}]{PhysRevA.88.053632}%
  \BibitemOpen
  \bibfield  {author} {\bibinfo {author} {\bibfnamefont {S.~P.}\ \bibnamefont
  {Rath}}\ and\ \bibinfo {author} {\bibfnamefont {R.}~\bibnamefont {Schmidt}},\
  }\href {https://doi.org/10.1103/PhysRevA.88.053632} {\bibfield  {journal}
  {\bibinfo  {journal} {Phys. Rev. A}\ }\textbf {\bibinfo {volume} {88}},\
  \bibinfo {pages} {053632} (\bibinfo {year} {2013})}\BibitemShut {NoStop}%
\bibitem [{\citenamefont {Li}\ and\ \citenamefont
  {Das~Sarma}(2014)}]{PhysRevA.90.013618}%
  \BibitemOpen
  \bibfield  {author} {\bibinfo {author} {\bibfnamefont {W.}~\bibnamefont
  {Li}}\ and\ \bibinfo {author} {\bibfnamefont {S.}~\bibnamefont {Das~Sarma}},\
  }\href {https://doi.org/10.1103/PhysRevA.90.013618} {\bibfield  {journal}
  {\bibinfo  {journal} {Phys. Rev. A}\ }\textbf {\bibinfo {volume} {90}},\
  \bibinfo {pages} {013618} (\bibinfo {year} {2014})}\BibitemShut {NoStop}%
\bibitem [{\citenamefont {Shashi}\ \emph {et~al.}(2014)\citenamefont {Shashi},
  \citenamefont {Grusdt}, \citenamefont {Abanin},\ and\ \citenamefont
  {Demler}}]{PhysRevA.89.053617}%
  \BibitemOpen
  \bibfield  {author} {\bibinfo {author} {\bibfnamefont {A.}~\bibnamefont
  {Shashi}}, \bibinfo {author} {\bibfnamefont {F.}~\bibnamefont {Grusdt}},
  \bibinfo {author} {\bibfnamefont {D.~A.}\ \bibnamefont {Abanin}},\ and\
  \bibinfo {author} {\bibfnamefont {E.}~\bibnamefont {Demler}},\ }\href
  {https://doi.org/10.1103/PhysRevA.89.053617} {\bibfield  {journal} {\bibinfo
  {journal} {Phys. Rev. A}\ }\textbf {\bibinfo {volume} {89}},\ \bibinfo
  {pages} {053617} (\bibinfo {year} {2014})}\BibitemShut {NoStop}%
\bibitem [{\citenamefont {Blinova}\ \emph {et~al.}(2013)\citenamefont
  {Blinova}, \citenamefont {Boshier},\ and\ \citenamefont
  {Timmermans}}]{PhysRevA.88.053610}%
  \BibitemOpen
  \bibfield  {author} {\bibinfo {author} {\bibfnamefont {A.~A.}\ \bibnamefont
  {Blinova}}, \bibinfo {author} {\bibfnamefont {M.~G.}\ \bibnamefont
  {Boshier}},\ and\ \bibinfo {author} {\bibfnamefont {E.}~\bibnamefont
  {Timmermans}},\ }\href {https://doi.org/10.1103/PhysRevA.88.053610}
  {\bibfield  {journal} {\bibinfo  {journal} {Phys. Rev. A}\ }\textbf {\bibinfo
  {volume} {88}},\ \bibinfo {pages} {053610} (\bibinfo {year}
  {2013})}\BibitemShut {NoStop}%
\bibitem [{\citenamefont {Ardila}\ and\ \citenamefont
  {Pohl}(2018)}]{Ardila_2018}%
  \BibitemOpen
  \bibfield  {author} {\bibinfo {author} {\bibfnamefont {L.~A.~P.}\
  \bibnamefont {Ardila}}\ and\ \bibinfo {author} {\bibfnamefont
  {T.}~\bibnamefont {Pohl}},\ }\href {https://doi.org/10.1088/1361-6455/aaf35e}
  {\bibfield  {journal} {\bibinfo  {journal} {J. Phys. B: At. Mol. Opt. Phys.}\
  }\textbf {\bibinfo {volume} {52}},\ \bibinfo {pages} {015004} (\bibinfo
  {year} {2018})}\BibitemShut {NoStop}%
\bibitem [{\citenamefont {Guenther}\ \emph {et~al.}(2018)\citenamefont
  {Guenther}, \citenamefont {Massignan}, \citenamefont {Lewenstein},\ and\
  \citenamefont {Bruun}}]{PhysRevLett.120.050405}%
  \BibitemOpen
  \bibfield  {author} {\bibinfo {author} {\bibfnamefont {N.-E.}\ \bibnamefont
  {Guenther}}, \bibinfo {author} {\bibfnamefont {P.}~\bibnamefont {Massignan}},
  \bibinfo {author} {\bibfnamefont {M.}~\bibnamefont {Lewenstein}},\ and\
  \bibinfo {author} {\bibfnamefont {G.~M.}\ \bibnamefont {Bruun}},\ }\href
  {https://doi.org/10.1103/PhysRevLett.120.050405} {\bibfield  {journal}
  {\bibinfo  {journal} {Phys. Rev. Lett.}\ }\textbf {\bibinfo {volume} {120}},\
  \bibinfo {pages} {050405} (\bibinfo {year} {2018})}\BibitemShut {NoStop}%
\bibitem [{\citenamefont {Ichmoukhamedov}\ and\ \citenamefont
  {Tempere}(2019)}]{PhysRevA.100.043605}%
  \BibitemOpen
  \bibfield  {author} {\bibinfo {author} {\bibfnamefont {T.}~\bibnamefont
  {Ichmoukhamedov}}\ and\ \bibinfo {author} {\bibfnamefont {J.}~\bibnamefont
  {Tempere}},\ }\href {https://doi.org/10.1103/PhysRevA.100.043605} {\bibfield
  {journal} {\bibinfo  {journal} {Phys. Rev. A}\ }\textbf {\bibinfo {volume}
  {100}},\ \bibinfo {pages} {043605} (\bibinfo {year} {2019})}\BibitemShut
  {NoStop}%
\bibitem [{\citenamefont {Field}\ \emph {et~al.}(2020)\citenamefont {Field},
  \citenamefont {Levinsen},\ and\ \citenamefont
  {Parish}}]{PhysRevA.101.013623}%
  \BibitemOpen
  \bibfield  {author} {\bibinfo {author} {\bibfnamefont {B.}~\bibnamefont
  {Field}}, \bibinfo {author} {\bibfnamefont {J.}~\bibnamefont {Levinsen}},\
  and\ \bibinfo {author} {\bibfnamefont {M.~M.}\ \bibnamefont {Parish}},\
  }\href {https://doi.org/10.1103/PhysRevA.101.013623} {\bibfield  {journal}
  {\bibinfo  {journal} {Phys. Rev. A}\ }\textbf {\bibinfo {volume} {101}},\
  \bibinfo {pages} {013623} (\bibinfo {year} {2020})}\BibitemShut {NoStop}%
\bibitem [{\citenamefont {Chin}\ \emph {et~al.}(2010)\citenamefont {Chin},
  \citenamefont {Grimm}, \citenamefont {Julienne},\ and\ \citenamefont
  {Tiesinga}}]{RevModPhys.82.1225}%
  \BibitemOpen
  \bibfield  {author} {\bibinfo {author} {\bibfnamefont {C.}~\bibnamefont
  {Chin}}, \bibinfo {author} {\bibfnamefont {R.}~\bibnamefont {Grimm}},
  \bibinfo {author} {\bibfnamefont {P.}~\bibnamefont {Julienne}},\ and\
  \bibinfo {author} {\bibfnamefont {E.}~\bibnamefont {Tiesinga}},\ }\href
  {https://doi.org/10.1103/RevModPhys.82.1225} {\bibfield  {journal} {\bibinfo
  {journal} {Rev. Mod. Phys.}\ }\textbf {\bibinfo {volume} {82}},\ \bibinfo
  {pages} {1225} (\bibinfo {year} {2010})}\BibitemShut {NoStop}%
\bibitem [{\citenamefont {Toyozawa}(1961)}]{10.1143/PTP.26.29}%
  \BibitemOpen
  \bibfield  {author} {\bibinfo {author} {\bibfnamefont {Y.}~\bibnamefont
  {Toyozawa}},\ }\href {https://doi.org/10.1143/PTP.26.29} {\bibfield
  {journal} {\bibinfo  {journal} {Prog. Theor. Phys.}\ }\textbf {\bibinfo
  {volume} {26}},\ \bibinfo {pages} {29} (\bibinfo {year} {1961})}\BibitemShut
  {NoStop}%
\bibitem [{\citenamefont {Sumi}\ and\ \citenamefont
  {Toyozawa}(1973)}]{1973137}%
  \BibitemOpen
  \bibfield  {author} {\bibinfo {author} {\bibfnamefont {A.}~\bibnamefont
  {Sumi}}\ and\ \bibinfo {author} {\bibfnamefont {Y.}~\bibnamefont
  {Toyozawa}},\ }\href {https://doi.org/10.1143/JPSJ.35.137} {\bibfield
  {journal} {\bibinfo  {journal} {J. Phys. Soc. Japan}\ }\textbf {\bibinfo
  {volume} {35}},\ \bibinfo {pages} {137} (\bibinfo {year} {1973})}\BibitemShut
  {NoStop}%
\bibitem [{\citenamefont {de~Gennes}(1960)}]{PhysRev.118.141}%
  \BibitemOpen
  \bibfield  {author} {\bibinfo {author} {\bibfnamefont {P.~G.}\ \bibnamefont
  {de~Gennes}},\ }\href {https://doi.org/10.1103/PhysRev.118.141} {\bibfield
  {journal} {\bibinfo  {journal} {Phys. Rev.}\ }\textbf {\bibinfo {volume}
  {118}},\ \bibinfo {pages} {141} (\bibinfo {year} {1960})}\BibitemShut
  {NoStop}%
\bibitem [{\citenamefont {Li}\ \emph {et~al.}(2015)\citenamefont {Li},
  \citenamefont {Zhu}, \citenamefont {He}, \citenamefont {Wang}, \citenamefont
  {Guo}, \citenamefont {Xu}, \citenamefont {Zhang},\ and\ \citenamefont
  {Wang}}]{PhysRevLett.114.255301}%
  \BibitemOpen
  \bibfield  {author} {\bibinfo {author} {\bibfnamefont {X.}~\bibnamefont
  {Li}}, \bibinfo {author} {\bibfnamefont {B.}~\bibnamefont {Zhu}}, \bibinfo
  {author} {\bibfnamefont {X.}~\bibnamefont {He}}, \bibinfo {author}
  {\bibfnamefont {F.}~\bibnamefont {Wang}}, \bibinfo {author} {\bibfnamefont
  {M.}~\bibnamefont {Guo}}, \bibinfo {author} {\bibfnamefont {Z.-F.}\
  \bibnamefont {Xu}}, \bibinfo {author} {\bibfnamefont {S.}~\bibnamefont
  {Zhang}},\ and\ \bibinfo {author} {\bibfnamefont {D.}~\bibnamefont {Wang}},\
  }\href {https://doi.org/10.1103/PhysRevLett.114.255301} {\bibfield  {journal}
  {\bibinfo  {journal} {Phys. Rev. Lett.}\ }\textbf {\bibinfo {volume} {114}},\
  \bibinfo {pages} {255301} (\bibinfo {year} {2015})}\BibitemShut {NoStop}%
\bibitem [{\citenamefont {Devreese}\ and\ \citenamefont
  {Alexandrov}(2009)}]{Devreese_2009}%
  \BibitemOpen
  \bibfield  {author} {\bibinfo {author} {\bibfnamefont {J.~T.}\ \bibnamefont
  {Devreese}}\ and\ \bibinfo {author} {\bibfnamefont {A.~S.}\ \bibnamefont
  {Alexandrov}},\ }\href {https://doi.org/10.1088/0034-4885/72/6/066501}
  {\bibfield  {journal} {\bibinfo  {journal} {Rep. Prog. Phys.}\ }\textbf
  {\bibinfo {volume} {72}},\ \bibinfo {pages} {066501} (\bibinfo {year}
  {2009})}\BibitemShut {NoStop}%
\bibitem [{\citenamefont {Mitra}\ \emph {et~al.}(1987)\citenamefont {Mitra},
  \citenamefont {Chatterjee},\ and\ \citenamefont
  {Mukhopadhyay}}]{MITRA198791}%
  \BibitemOpen
  \bibfield  {author} {\bibinfo {author} {\bibfnamefont {T.}~\bibnamefont
  {Mitra}}, \bibinfo {author} {\bibfnamefont {A.}~\bibnamefont {Chatterjee}},\
  and\ \bibinfo {author} {\bibfnamefont {S.}~\bibnamefont {Mukhopadhyay}},\
  }\href {https://doi.org/https://doi.org/10.1016/0370-1573(87)90087-1}
  {\bibfield  {journal} {\bibinfo  {journal} {Phys. Rep.}\ }\textbf {\bibinfo
  {volume} {153}},\ \bibinfo {pages} {91 } (\bibinfo {year}
  {1987})}\BibitemShut {NoStop}%
\bibitem [{\citenamefont {Ho}(1998)}]{PhysRevLett.81.742}%
  \BibitemOpen
  \bibfield  {author} {\bibinfo {author} {\bibfnamefont {T.-L.}\ \bibnamefont
  {Ho}},\ }\href {https://doi.org/10.1103/PhysRevLett.81.742} {\bibfield
  {journal} {\bibinfo  {journal} {Phys. Rev. Lett.}\ }\textbf {\bibinfo
  {volume} {81}},\ \bibinfo {pages} {742} (\bibinfo {year} {1998})}\BibitemShut
  {NoStop}%
\bibitem [{\citenamefont {Xu}\ \emph {et~al.}(2012)\citenamefont {Xu},
  \citenamefont {Wang},\ and\ \citenamefont {You}}]{PhysRevA.86.013632}%
  \BibitemOpen
  \bibfield  {author} {\bibinfo {author} {\bibfnamefont {Z.~F.}\ \bibnamefont
  {Xu}}, \bibinfo {author} {\bibfnamefont {D.~J.}\ \bibnamefont {Wang}},\ and\
  \bibinfo {author} {\bibfnamefont {L.}~\bibnamefont {You}},\ }\href
  {https://doi.org/10.1103/PhysRevA.86.013632} {\bibfield  {journal} {\bibinfo
  {journal} {Phys. Rev. A}\ }\textbf {\bibinfo {volume} {86}},\ \bibinfo
  {pages} {013632} (\bibinfo {year} {2012})}\BibitemShut {NoStop}%
\bibitem [{\citenamefont {Bogolyubov}(1947)}]{Bogolyubov:1947zz}%
  \BibitemOpen
  \bibfield  {author} {\bibinfo {author} {\bibfnamefont {N.}~\bibnamefont
  {Bogolyubov}},\ }\href@noop {} {\bibfield  {journal} {\bibinfo  {journal} {J.
  Phys. (USSR)}\ }\textbf {\bibinfo {volume} {11}},\ \bibinfo {pages} {23}
  (\bibinfo {year} {1947})}\BibitemShut {NoStop}%
\bibitem [{\citenamefont {Kawaguchi}\ and\ \citenamefont
  {Ueda}(2012)}]{KAWAGUCHI2012253}%
  \BibitemOpen
  \bibfield  {author} {\bibinfo {author} {\bibfnamefont {Y.}~\bibnamefont
  {Kawaguchi}}\ and\ \bibinfo {author} {\bibfnamefont {M.}~\bibnamefont
  {Ueda}},\ }\href
  {https://doi.org/https://doi.org/10.1016/j.physrep.2012.07.005} {\bibfield
  {journal} {\bibinfo  {journal} {Phys. Rep.}\ }\textbf {\bibinfo {volume}
  {520}},\ \bibinfo {pages} {253 } (\bibinfo {year} {2012})}\BibitemShut
  {NoStop}%
\bibitem [{\citenamefont {Fröhlich}\ \emph {et~al.}(1950)\citenamefont
  {Fröhlich}, \citenamefont {Pelzer},\ and\ \citenamefont
  {Zienau}}]{doi:10.1080/14786445008521794}%
  \BibitemOpen
  \bibfield  {author} {\bibinfo {author} {\bibfnamefont {H.}~\bibnamefont
  {Fröhlich}}, \bibinfo {author} {\bibfnamefont {H.}~\bibnamefont {Pelzer}},\
  and\ \bibinfo {author} {\bibfnamefont {S.}~\bibnamefont {Zienau}},\ }\href
  {https://doi.org/10.1080/14786445008521794} {\bibfield  {journal} {\bibinfo
  {journal} {Phil. Mag.}\ }\textbf {\bibinfo {volume} {41}},\ \bibinfo {pages}
  {221} (\bibinfo {year} {1950})}\BibitemShut {NoStop}%
\bibitem [{\citenamefont {Feynman}(2018)}]{feynman2018statistical}%
  \BibitemOpen
  \bibfield  {author} {\bibinfo {author} {\bibfnamefont {R.}~\bibnamefont
  {Feynman}},\ }\href {https://books.google.com/books?id=1a1SDwAAQBAJ} {\emph
  {\bibinfo {title} {Statistical Mechanics: A Set Of Lectures}}}\ (\bibinfo
  {publisher} {CRC Press},\ \bibinfo {year} {2018})\BibitemShut {NoStop}%
\bibitem [{\citenamefont {Lee}\ \emph {et~al.}(1953)\citenamefont {Lee},
  \citenamefont {Low},\ and\ \citenamefont {Pines}}]{PhysRev.90.297}%
  \BibitemOpen
  \bibfield  {author} {\bibinfo {author} {\bibfnamefont {T.~D.}\ \bibnamefont
  {Lee}}, \bibinfo {author} {\bibfnamefont {F.~E.}\ \bibnamefont {Low}},\ and\
  \bibinfo {author} {\bibfnamefont {D.}~\bibnamefont {Pines}},\ }\href
  {https://doi.org/10.1103/PhysRev.90.297} {\bibfield  {journal} {\bibinfo
  {journal} {Phys. Rev.}\ }\textbf {\bibinfo {volume} {90}},\ \bibinfo {pages}
  {297} (\bibinfo {year} {1953})}\BibitemShut {NoStop}%
\bibitem [{\citenamefont {van Kempen}\ \emph {et~al.}(2002)\citenamefont {van
  Kempen}, \citenamefont {Kokkelmans}, \citenamefont {Heinzen},\ and\
  \citenamefont {Verhaar}}]{PhysRevLett.88.093201}%
  \BibitemOpen
  \bibfield  {author} {\bibinfo {author} {\bibfnamefont {E.~G.~M.}\
  \bibnamefont {van Kempen}}, \bibinfo {author} {\bibfnamefont {S.~J. J.
  M.~F.}\ \bibnamefont {Kokkelmans}}, \bibinfo {author} {\bibfnamefont {D.~J.}\
  \bibnamefont {Heinzen}},\ and\ \bibinfo {author} {\bibfnamefont {B.~J.}\
  \bibnamefont {Verhaar}},\ }\href
  {https://doi.org/10.1103/PhysRevLett.88.093201} {\bibfield  {journal}
  {\bibinfo  {journal} {Phys. Rev. Lett.}\ }\textbf {\bibinfo {volume} {88}},\
  \bibinfo {pages} {093201} (\bibinfo {year} {2002})}\BibitemShut {NoStop}%
\bibitem [{\citenamefont {Landau}(1933)}]{Landau:1933iwn}%
  \BibitemOpen
  \bibfield  {author} {\bibinfo {author} {\bibfnamefont {L.}~\bibnamefont
  {Landau}},\ }\href@noop {} {\bibfield  {journal} {\bibinfo  {journal} {Phys.
  Z. Sowjetunion}\ }\textbf {\bibinfo {volume} {3}},\ \bibinfo {pages} {644}
  (\bibinfo {year} {1933})}\BibitemShut {NoStop}%
\bibitem [{\citenamefont {Landau}\ and\ \citenamefont {Pekar}(1948)}]{Pekar}%
  \BibitemOpen
  \bibfield  {author} {\bibinfo {author} {\bibfnamefont {L.}~\bibnamefont
  {Landau}}\ and\ \bibinfo {author} {\bibfnamefont {S.}~\bibnamefont {Pekar}},\
  }\href {http://archive.ujp.bitp.kiev.ua/files/journals/53/si/53SI15p.pdf}
  {\bibfield  {journal} {\bibinfo  {journal} {Zh. Eksp. Teor. Fiz.}\ }\textbf
  {\bibinfo {volume} {18}},\ \bibinfo {pages} {419} (\bibinfo {year}
  {1948})}\BibitemShut {NoStop}%
\bibitem [{\citenamefont {Providencia}\ \emph {et~al.}(1975)\citenamefont
  {Providencia}, \citenamefont {Ruivo},\ and\ \citenamefont
  {Sousa}}]{DAPROVIDENCIA1975366}%
  \BibitemOpen
  \bibfield  {author} {\bibinfo {author} {\bibfnamefont {J.}~\bibnamefont
  {Providencia}}, \bibinfo {author} {\bibfnamefont {M.}~\bibnamefont {Ruivo}},\
  and\ \bibinfo {author} {\bibfnamefont {C.}~\bibnamefont {Sousa}},\ }\href
  {https://doi.org/https://doi.org/10.1016/0003-4916(75)90226-2} {\bibfield
  {journal} {\bibinfo  {journal} {Ann. Phys.}\ }\textbf {\bibinfo {volume}
  {91}},\ \bibinfo {pages} {366 } (\bibinfo {year} {1975})}\BibitemShut
  {NoStop}%
\bibitem [{\citenamefont {Peeters}\ and\ \citenamefont
  {Devreese}(1985)}]{PhysRevB.32.3515}%
  \BibitemOpen
  \bibfield  {author} {\bibinfo {author} {\bibfnamefont {F.~M.}\ \bibnamefont
  {Peeters}}\ and\ \bibinfo {author} {\bibfnamefont {J.~T.}\ \bibnamefont
  {Devreese}},\ }\href {https://doi.org/10.1103/PhysRevB.32.3515} {\bibfield
  {journal} {\bibinfo  {journal} {Phys. Rev. B}\ }\textbf {\bibinfo {volume}
  {32}},\ \bibinfo {pages} {3515} (\bibinfo {year} {1985})}\BibitemShut
  {NoStop}%
\bibitem [{\citenamefont {Pethick}\ and\ \citenamefont
  {Smith}(2008)}]{pethick_smith_2008}%
  \BibitemOpen
  \bibfield  {author} {\bibinfo {author} {\bibfnamefont {C.~J.}\ \bibnamefont
  {Pethick}}\ and\ \bibinfo {author} {\bibfnamefont {H.}~\bibnamefont
  {Smith}},\ }\href {https://doi.org/10.1017/CBO9780511802850} {\emph {\bibinfo
  {title} {Bose–Einstein Condensation in Dilute Gases}}},\ \bibinfo {edition}
  {2nd}\ ed.\ (\bibinfo  {publisher} {Cambridge University Press},\ \bibinfo
  {year} {2008})\BibitemShut {NoStop}%
\bibitem [{\citenamefont {Papoular}\ \emph {et~al.}(2010)\citenamefont
  {Papoular}, \citenamefont {Shlyapnikov},\ and\ \citenamefont
  {Dalibard}}]{PhysRevA.81.041603}%
  \BibitemOpen
  \bibfield  {author} {\bibinfo {author} {\bibfnamefont {D.~J.}\ \bibnamefont
  {Papoular}}, \bibinfo {author} {\bibfnamefont {G.~V.}\ \bibnamefont
  {Shlyapnikov}},\ and\ \bibinfo {author} {\bibfnamefont {J.}~\bibnamefont
  {Dalibard}},\ }\href {https://doi.org/10.1103/PhysRevA.81.041603} {\bibfield
  {journal} {\bibinfo  {journal} {Phys. Rev. A}\ }\textbf {\bibinfo {volume}
  {81}},\ \bibinfo {pages} {041603} (\bibinfo {year} {2010})}\BibitemShut
  {NoStop}%
\end{thebibliography}%
\end{document}